\documentclass[12pt]{article}
\usepackage{epsfig}
\title{Atmospheric Neutrino Fluxes\thanks{Research
supported in part by the U.S. Department of Energy under DE-FG02 91ER40626.}
}
\author{Thomas K. Gaisser\\
Bartol Research Institute, University of Delaware\\
Newark, DE 19716 USA}
\begin{document}
\maketitle
\begin{abstract}
Starting with an historical review, I summarize the status of
calculations of the flux of atmospheric neutrinos and how they
compare to the measurements.
\end{abstract}

\section{Introduction}

When cosmic-ray protons and nuclei enter the atmosphere, 
they interact and produce all kinds of secondary particles, 
which in turn interact or decay or propagate to the ground, 
depending on their intrinsic properties and energies.  
In the GeV energy range, the most abundant particles at the 
ground are the neutrinos. At production, there are
 approximately twice as many muon neutrinos as 
electron neutrinos.  This characteristic ratio arises from the 
pion-muon-electron decay chain in which a muon and one muon-neutrino 
(or anti-neutrino) are produced when 
the charged pion decays, while the subsequent muon decay produces 
an electron neutrino as well as another muon neutrino.  

In 1960 Markov~\cite{Markov} suggested using Cherenkov light 
in a lake or the deep ocean
to do neutrino physics with the atmospheric neutrino beam;
in particular, to investigate the question whether electron and muon
neutrinos are distinct species.  About the same time Greisen~\cite{Greisen}
described a proposal to search for astrophysical sources of neutrinos
with a water Cherenkov detector in a deep mine.
Atmospheric neutrinos were, however, first detected with electronic detectors--as 
horizontal neutrino-induced muons in a mine in South Africa~\cite{CW} and
in an iron calorimeter in a deep mine
in the Kolar Gold Fields of India~\cite{KGF}.  More than a decade
passed before a large-scale effort to instrument a large body of 
water was undertaken by the DUMAND Project~\cite{DUMAND}.

Atmospheric neutrino fluxes were first calculated in the 1960's.  One
approach~\cite{MZ} is to infer the neutrino fluxes from measurements
of closely related muons.  More recently this approach has been
refined to take account of effects of the geomagnetic field and of the
separate contributions of pions and kaons~\cite{Perkins}.  Most later
calculations follow Refs.~\cite{ZK,ZK2} and calculate both
the muon~\cite{ZK} and the neutrino
fluxes~\cite{ZK2} starting from the primary cosmic-ray spectrum at the top of
the atmosphere. 

With the advent in the 1980's of large 
underground detectors to search for proton decay, 
it became possible to measure increasingly large samples 
of events induced by neutrinos of both types 
(and from all directions because the Earth is 
transparent to neutrinos).  Such events were of interest primarily
 as the background for proton decay.  Both water Cherenkov 
detectors~\cite{IMB,Kam} and segmented iron 
calorimeters~\cite{KGF,Nusex,Frejus,Soudan} were used.

Evidence 
gradually accumulated that the ratio of the two neutrino flavors 
was different from the expected value of two.  The first hint came
from the IMB experiment~\cite{IMBstopmu}, which reported too few muon decays
compared to what was expected from from interactions
of $\nu_\mu$ inside their detector.  The IMB experiment later
concluded, however, that their measurement of the ratio of stopping to throughgoing
muons was inconsistent with oscillations~\cite{IMBPRL}.
The Kamiokande experiment~\cite{Kam88}
found an anomalously low ratio of $\nu_\mu\,/\,\nu_e$ and suggested 
neutrino oscillations as a possible explanation.  As the IMB experiment
accumulated more data~\cite{IMBlater}, they too found a persistently low 
$\nu_\mu\,/\,\nu_e$ flavor ratio.  Meanwhile, however, the Frejus experiment~\cite{Frejus}
measured a ratio consistent with the expectation.
After some six years of running with the shielded Kamiokande II-III detector,
enough data was accumulated to  
see a suggestion of the pathlength dependence expected from oscillations
with parameters such that upward $\nu_\mu$ in the multi-GeV energy region 
oscillate while downward neutrinos do not~\cite{Kam94}.

By measuring the directions and energies of thousands of neutrinos 
of both flavors in the past decade, the Super-Kamiokande 
Collaboration has shown that the anomalous ratio 
is a consequence of neutrino flavor oscillations during their 
propagation from the atmosphere to the detector~\cite{SK98}. 
 They have measured the oscillations in the muon neutrino--tau 
neutrino sector~\cite{SKA} by fitting the energy and 
angular dependence and other observed features of their data,
while ruling out oscillation to sterile neutrinos at 99\% 
confidence level.  It is now possible to demonstrate the
$L/E$ pathlength/energy dependence expected for oscillations~\cite{SKB}.
Best fit parameters in a two-flavor approximation
($\nu_\mu\,\leftrightarrow\,\nu_\tau$) are
$\delta m^2\,=\,2.1\times 10^{-3}$~eV$^2$ and maximal mixing~\cite{SK1489}.
Results from MACRO~\cite{MACRO} and Soudan~\cite{Soudan}
are consistent with these parameters.  
Measurements of atmospheric neutrinos are reviewed in 
Refs.~\cite{KT,Jung,MGoodman}.

With the parallel discovery~\cite{Sno1,SKsolar,Sno2} of oscillations of solar 
electron-neutrinos in a different region of parameter space, 
a pattern of mixing among the three neutrino types 
(electron, muon and tau) is beginning to emerge in which 
two mixing angles are large while one is smaller 
(though not yet measured).  Eventually the nature of the full 
neutrino mass matrix will be addressed with long-baseline 
accelerator neutrino beams by experiments now 
under construction or planned.  This will take some time, 
however.  Accordingly, there is interest in refining
calculations of the atmospheric neutrino beam to 
the point where effects of sub-dominant mixing 
may be resolved.  To take full advantage of the power of the 
Super-Kamiokande experiment requires reducing 
uncertainties in the absolute normalization of the neutrino 
flux and in the ratios of electron to muon neutrinos 
and neutrinos to anti-neutrinos.  The normalization depends 
on the primary cosmic-ray intensity and on the 
details of production of pions and kaons in cosmic-ray 
interactions in the atmosphere.  The ratios depend 
primarily on pion and kaon production.  This paper reviews 
the present level of uncertainties. 

\section{Calculation}

The neutrino flux is a convolution of the primary cosmic-ray spectrum,
modified by geomagnetic cutoffs, with the yield of neutrinos per incident
cosmic-ray.  Schematically,
\begin{equation}
\phi_\nu \;=\;\phi_p\,\otimes\,R_p\,\otimes\,Y_{p\rightarrow\nu} 
 \;+\;\sum_A\,\phi_A\,\otimes\,R_A\,\otimes\,Y_{A\rightarrow\nu}.
\label{convolution}
\end{equation}
The first term represents the contribution of free protons and the second
term the contribution of nucleons bound in nuclei.  The two need to be
calculated separately even in the superposition approximation in which
it is assumed that bound nucleons interact independently as if they were
free.  This is because the geomagnetic cutoff depends on rigidity
(total momentum per unit charge) whereas pion production depends on 
energy per nucleon.  In each term the factor $\phi(E)$ represents
the primary cosmic-ray spectrum as a function of energy per nucleon.
$R(E,\theta_\oplus,\phi_\oplus,\theta,\phi)$ 
indicates the cutoff rigidity (more exactly the transmission
probability) which depends on latitude and longitude 
and on the local zenith and
azimuth of the arrival direction of each primary cosmic ray.  
The yield $Y(E_A,E_\nu)$
gives the number of neutrinos with energy $E_\nu$ produced per
primary of energy $E_A$ and is obtained by calculating the
cosmic-ray induced cascades in the atmosphere.  A complete review
of the calculation of the flux of atmospheric neutrinos is given in
Ref.~\cite{GH}.  In this paper I illustrate the considerations
involved in calculating the production spectrum of atmospheric
neutrinos by comparing results of three recent 
calculations~\cite{FLUKA,Hondaetal,Barretal}.

All these calculations are three-dimensional, taking account of the
deviations of the neutrinos from the directions of the primaries that
produce them.  In addition, the calculations of Refs.~\cite{Hondaetal}
and~\cite{Barretal} also account for the effect bending of muons in
the geomagnetic field before they decay.  There are several other
fully three-dimensional calculations, including the recent
papers of Refs.~\cite{Wentz,Liu,Favier} which appeared after the
review~\cite{GH}.  The challenge of the full three-dimensional
calculation is indicated by the five arguments of the cutoff 
function above.  Cascades must be generated for relevant primary
energies all over the globe ($\theta_\oplus,\phi_\oplus$)
taking account of the cutoff for primaries from all directions
($\theta,\phi$).  Since most of the created neutrinos miss
the detector, the calculation is highly inefficient.  Although
three-dimensional aspects of a calculation are technically challenging,
they are not the most important sources of uncertainty in the
calculated fluxes of atmospheric neutrinos.  The biggest
uncertainties come from the primary cosmic-ray
spectrum and the treatment of hadronic interactions.

\subsection{Primary spectrum}

\begin{figure}[htb]
\includegraphics[width=14cm]{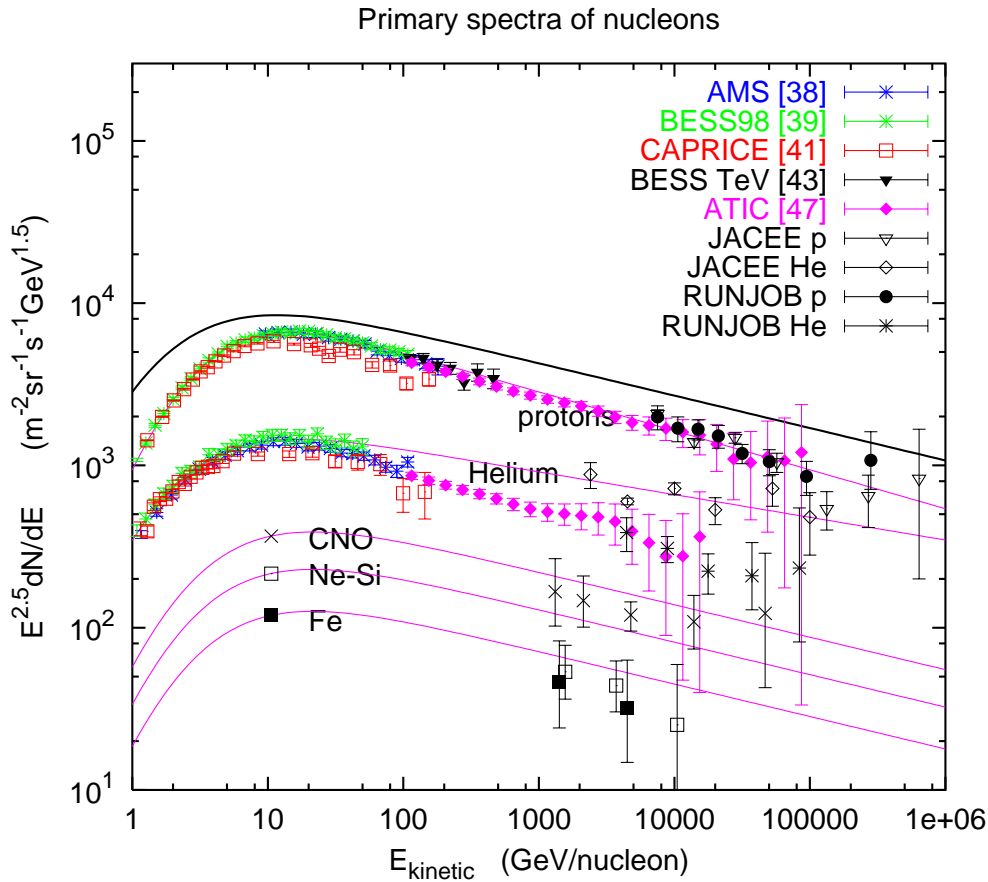}
\caption{The flux of nucleons.  The heavy black line shows the
numerical form of Eq.~\ref{uncorrelated}.  The lighter lines show
extrapolations of fits~\protect\cite{Hamburg} to measurements
of protons, helium and three heavier groups below 100 GeV/nucleon.
}
\label{allnucleon_t9}
\end{figure}

In the past decade new measurements of the primary spectrum with
magnetic spectrometers have improved our knowledge of the primary
spectrum.  In particular, the AMS~\cite{AMSp} and BESS~\cite{BESS}
detectors give results for the proton spectra up to 100 GeV
that agree with each other within 5\%.  As a consequence, current
calculations of atmospheric neutrinos are using fits to the
primary spectrum in which these data have been given priority.
An example of such a fit~\cite{Hamburg} is shown by the thin
lines in Fig.~\ref{allnucleon_t9}.  Here the various nuclear
groups are plotted as nucleons per GeV/nucleon, which is the
quantity most directly related to secondary particle production.
Free protons make up about 70\% of the flux of all nucleons,
helium about 20\%, and heavier nuclei the rest.
Another recent series of spectrometer measurements (CAPRICE~\cite{CAPRICE})
continues to give
results for protons about 15\% lower than BESS and AMS, which
can be taken as an indication of the uncertainty in the absolute
normalization of this component of the primary spectrum.  It should
also be noted that BESS~\cite{BESS} and AMS~\cite{AMSHe}
results for helium do not agree with each other as well as for protons.

Measurements with spectrometers presently extend up to about 500 GeV
only~\cite{BESSTeV}.  Higher energy data are from balloon-borne
ionization calorimeters, which sample
the fraction of energy deposited in electromagnetic cascades
inside the calorimeter.  Systematic errors in assigning
primary energy are larger as a consequence.  
Data from JACEE~\cite{JACEE} and 
RUNJOB~\cite{RUNJOB1,RUNJOB2} are included in Fig.~\ref{allnucleon_t9}.
Preliminary data from the ATIC experiment~\cite{ATIC} fill
in the gap in the 1--100 TeV region.  The ATIC
data for protons appear consistent with all previous data.
Their helium data, if confirmed, indicate a preference
for the lower RUNJOB normalization for helium.

\begin{figure}[htb]
\includegraphics[width=14cm]{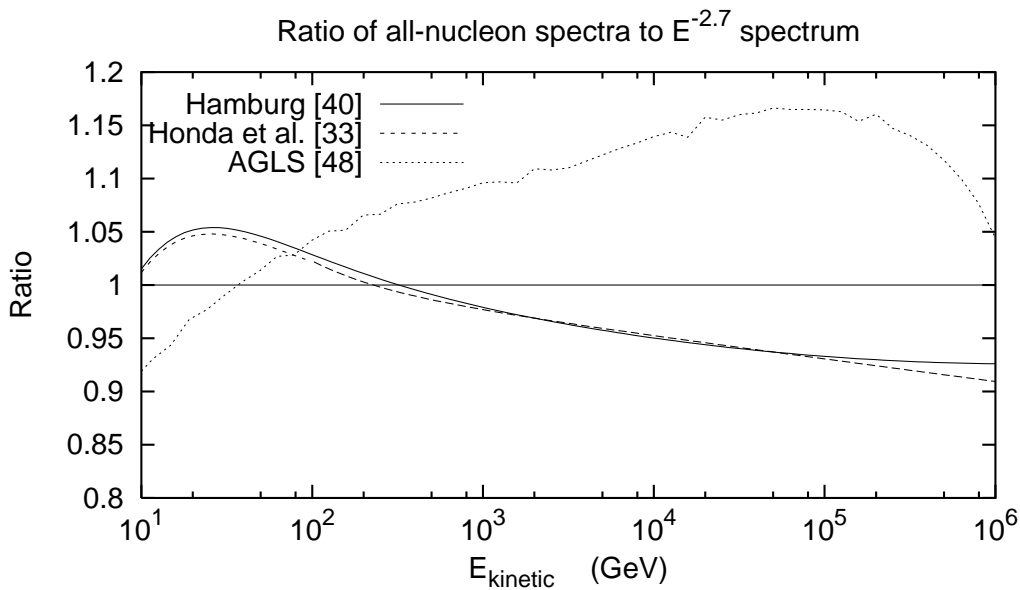}
\caption{Ratio of the all-nucleon fluxes used in several
calculations to the power-law form of Eq.~\protect\ref{uncorrelated}.
}
\label{allnucleon-ratios}
\end{figure}

The heavy solid line in Fig.~\ref{allnucleon_t9} is a simple 
power-law approximation to the spectrum of all nucleons,
\begin{equation}
\phi_N(E)\;=\;1.7\,{{\rm nucleons}\over cm^2 s\,sr\,GeV}\times 
\left({1\over E}\right)^{2.7},
\label{uncorrelated}
\end{equation}
where $E$ is total energy per nucleon.
Fig.~\ref{allnucleon-ratios} displays the all-nucleon spectra used to
calculate the neutrino flux as a ratio to the
power-law form of Eq.~\ref{uncorrelated}.  Refs.~\cite{FLUKA,Barretal}
use the fits of Ref.~\cite{Hamburg}, while Ref.~\cite{Hondaetal} use
slightly different parameters for the heavy components at low energy
and a slightly flatter power-law extrapolation (-2.71) above $100$~GeV.
Nevertheless, 
the primary all-nucleon spectrum of the three calculations~\cite{FLUKA,Hondaetal,Barretal} 
are the same to within about 1\%.
Moreover, simple power law
approximation~\ref{uncorrelated} to the all-nucleon spectrum
represents these fits 
to better than 10\% over the whole energy region.  For reference, 
Fig.~\ref{allnucleon-ratios} also shows the primary spectrum used in an
earlier calculation~\cite{AGLS}, which is somewhat lower than the new
fits below $100$~GeV and significantly higher at high energy.

\begin{figure}[htb]
\includegraphics[width=7.5cm]{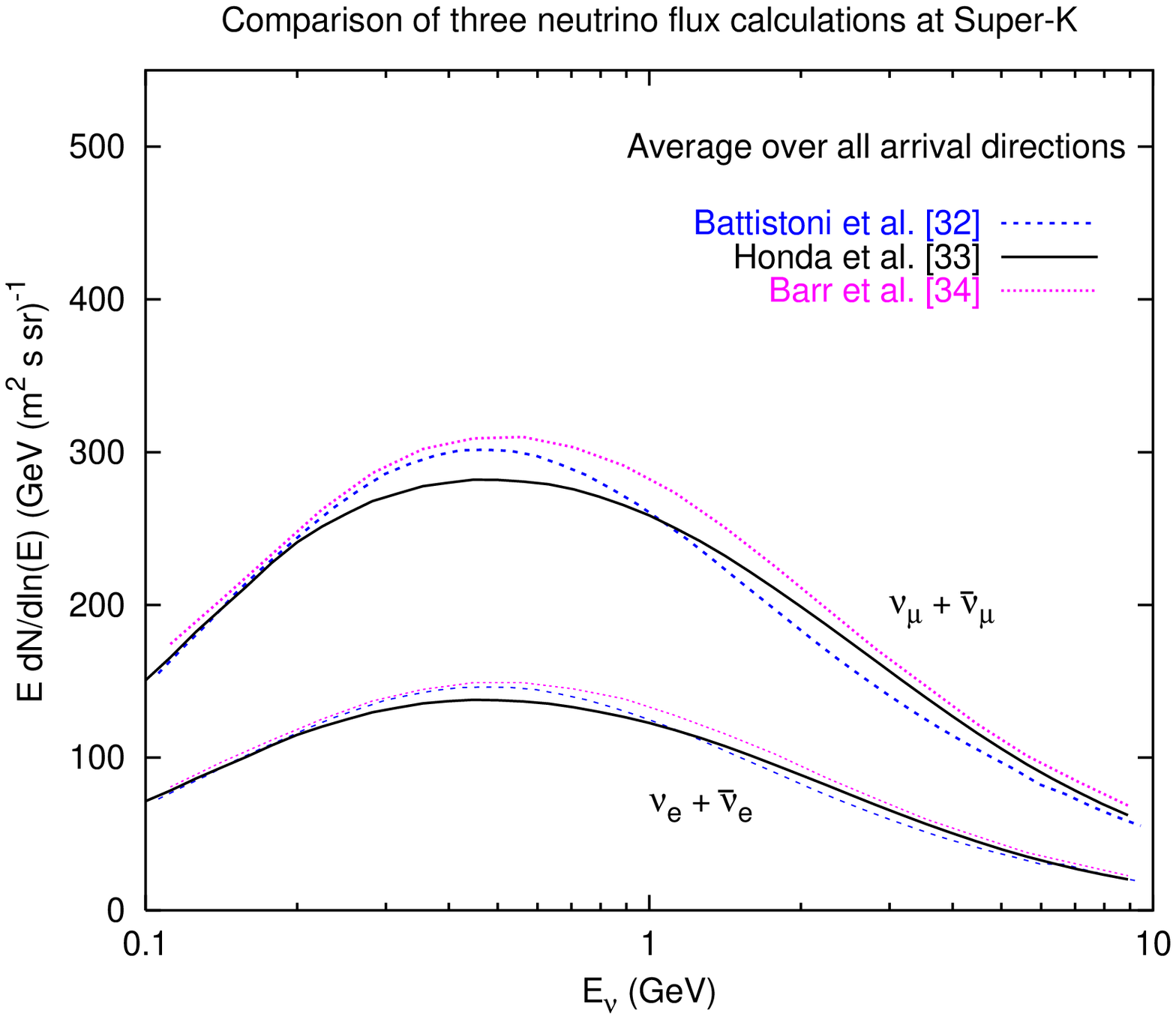}
\includegraphics[width=7.5cm]{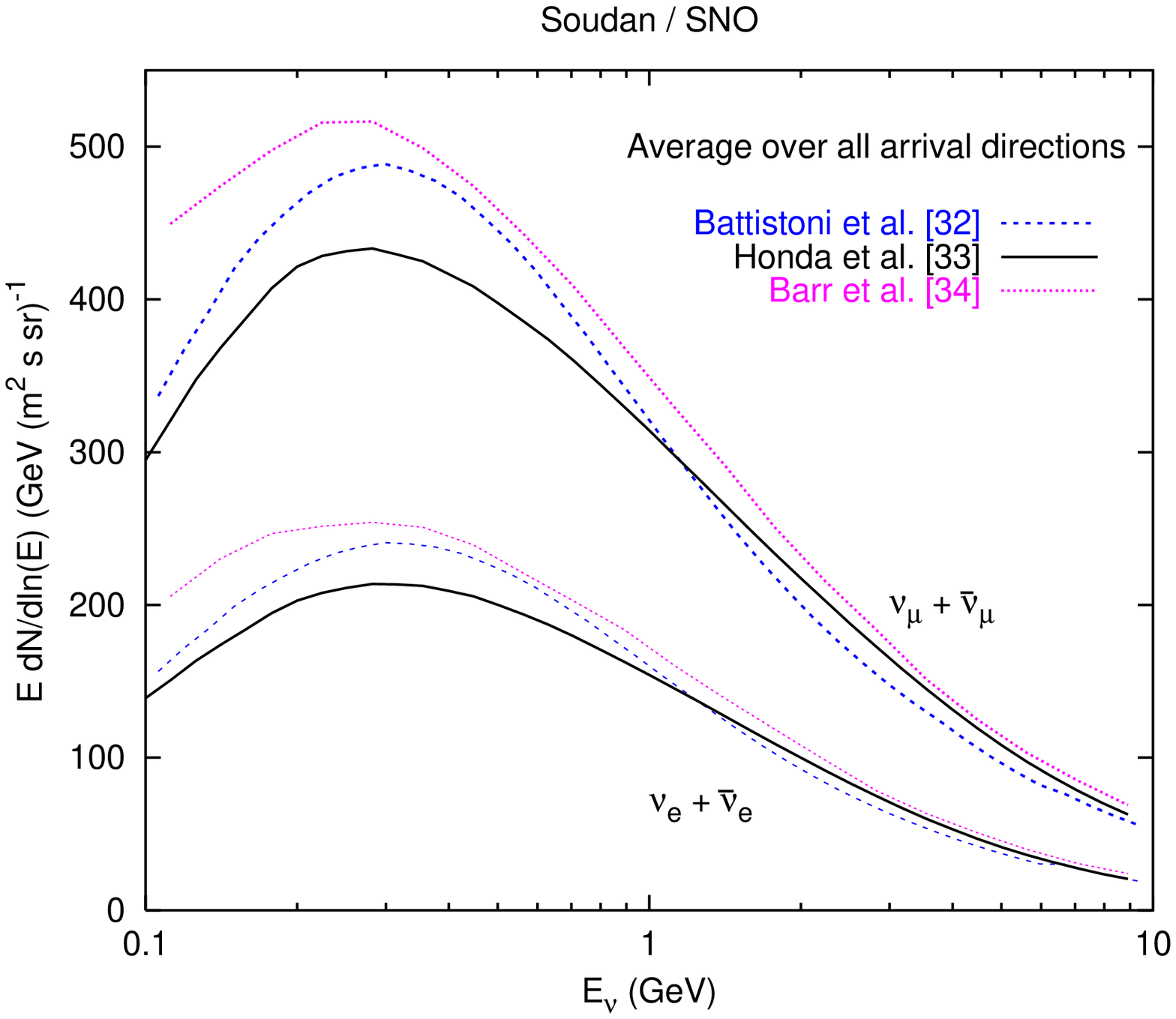}
\caption{The production spectrum of $\nu_\mu+\bar{\nu}_\mu$
and $\nu_e+\bar{\nu}_e$ integrated over the atmosphere and
averaged over all directions at Super-K (left) and Soudan or SNO (right).
}
\label{neutrinoflux}
\end{figure}

\subsection{Treatment of hadronic interactions}

The representations of hadronic interactions used in Refs.~\cite{FLUKA,Hondaetal,Barretal}
are completely independent.  As a consequence, it is not surprising that they
give different results.  Because the assumptions about primary spectrum
are essentially identical, differences among the results of the three
calculations reflect the level of uncertainty due to treatment of hadronic
interactions.  For low energies the fluxes depend also on the geomagnetic
cutoffs, so comparison must be made for the same location.
Figure~\ref{neutrinoflux}
shows the atmospheric neutrino fluxes from the three 
calculations~\cite{FLUKA,Hondaetal,Barretal}.
Differences among the calculations are at the level of 10\%
at Super-K.
The differences are somewhat larger for
the sites at high geomagnetic latitude (where the cutoff
for downward primaries is negligible), which probably indicates
that the low-energy hadronic interactions are less well 
understood.

\begin{figure}[htb]
\includegraphics[width=7.5cm]{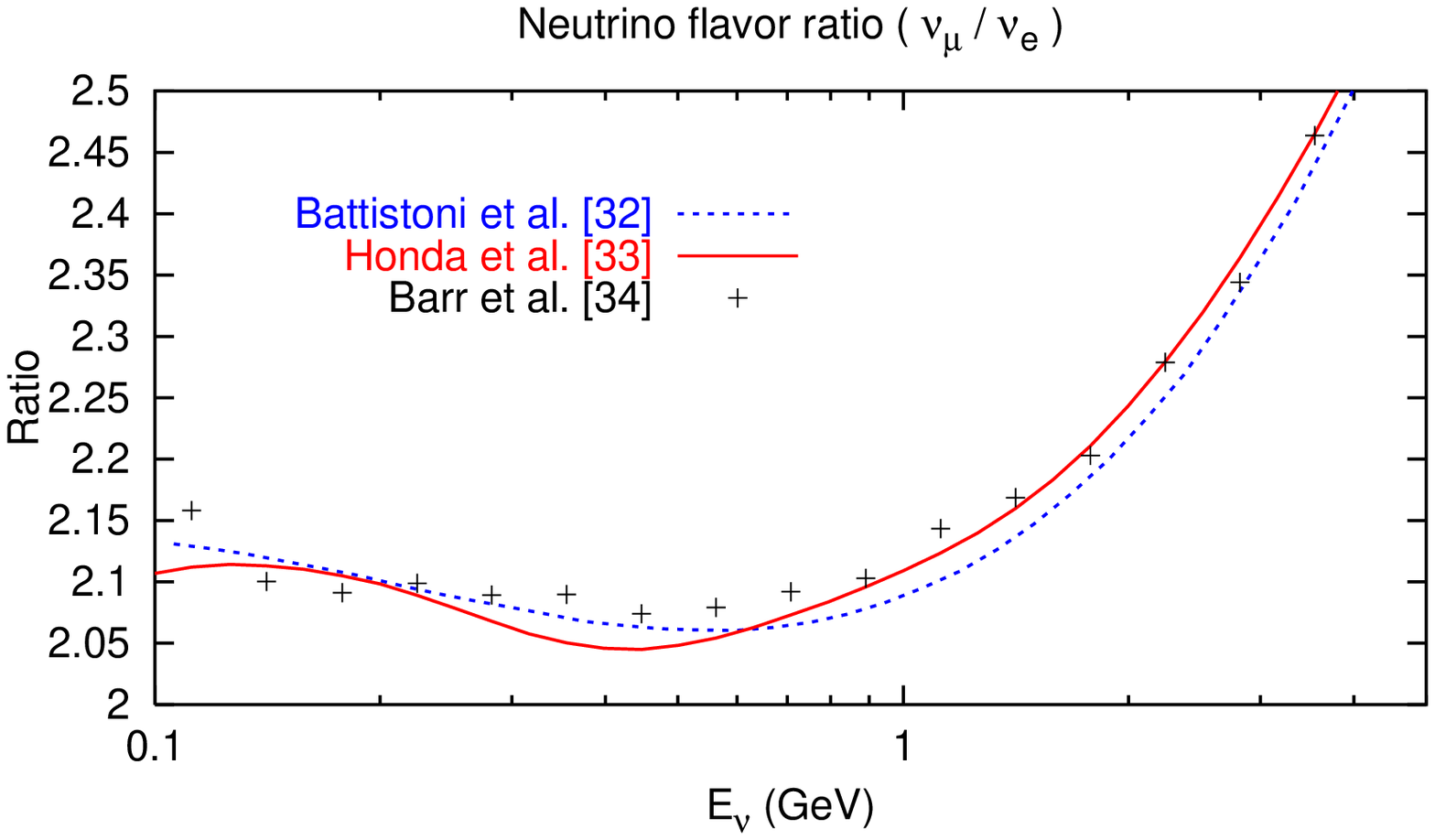}
\includegraphics[width=7.5cm]{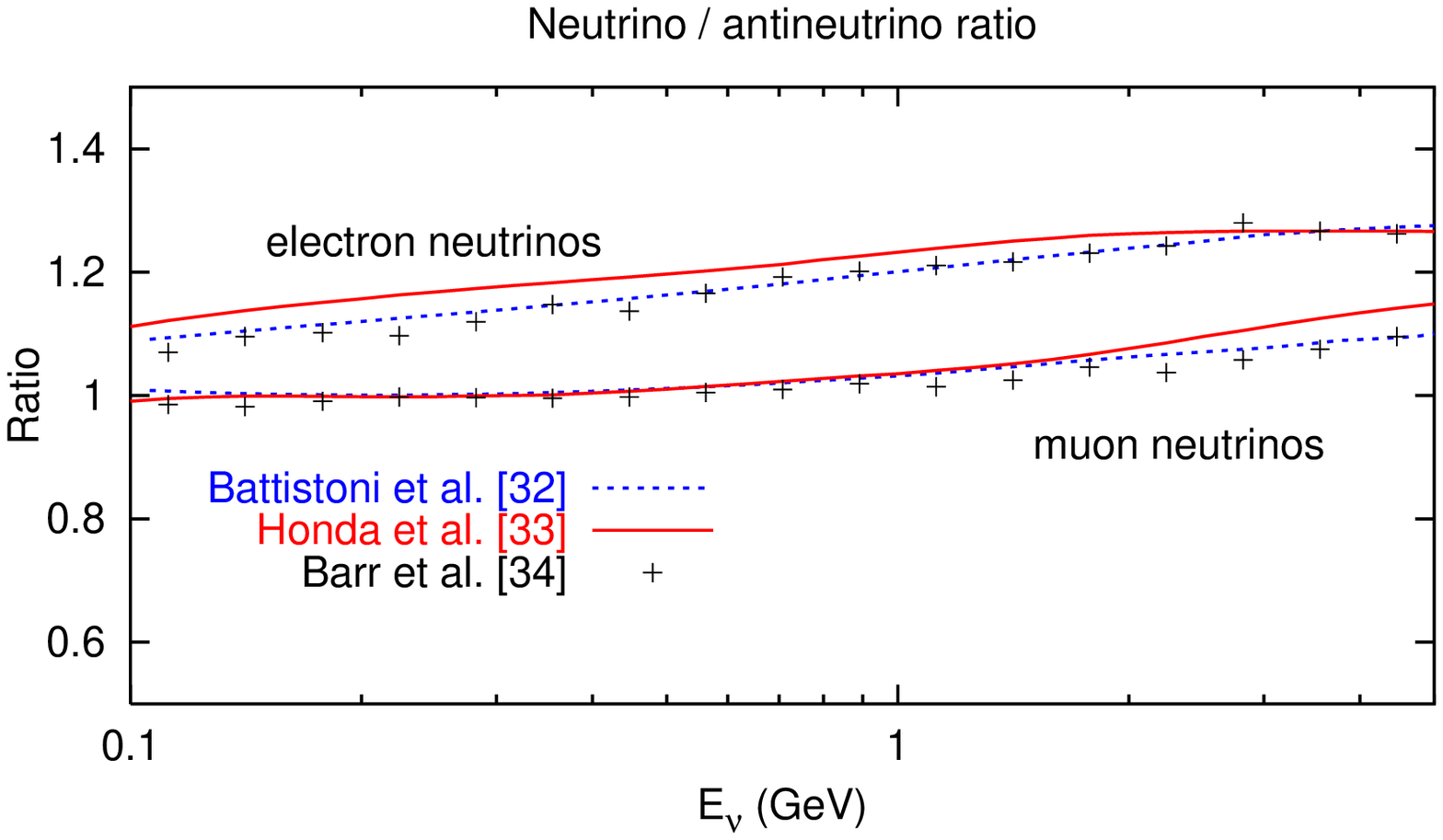}
\caption{The ratio $\nu_\mu+\bar{\nu}_\mu$ to
$\nu_e+\bar{\nu}_e$ (left panel) and $\nu /\bar{\nu}$ (right panel) at Super-K.  
The production spectra are integrated over 
the atmosphere neglecting oscillations and
averaged over all directions before taking the ratio.
}
\label{nuratios}
\end{figure}

The flavor ratios for the three calculations are shown for the low-energy
region in Fig.~\ref{nuratios}, along with the neutrino/antineutrino 
ratios.
Differences are at the level of 3\% in this energy range for
the flavor ratios and 5\% for the $\nu\,/\,\bar{\nu}$ ratios.

\begin{figure}[htb]
\includegraphics[width=8cm]{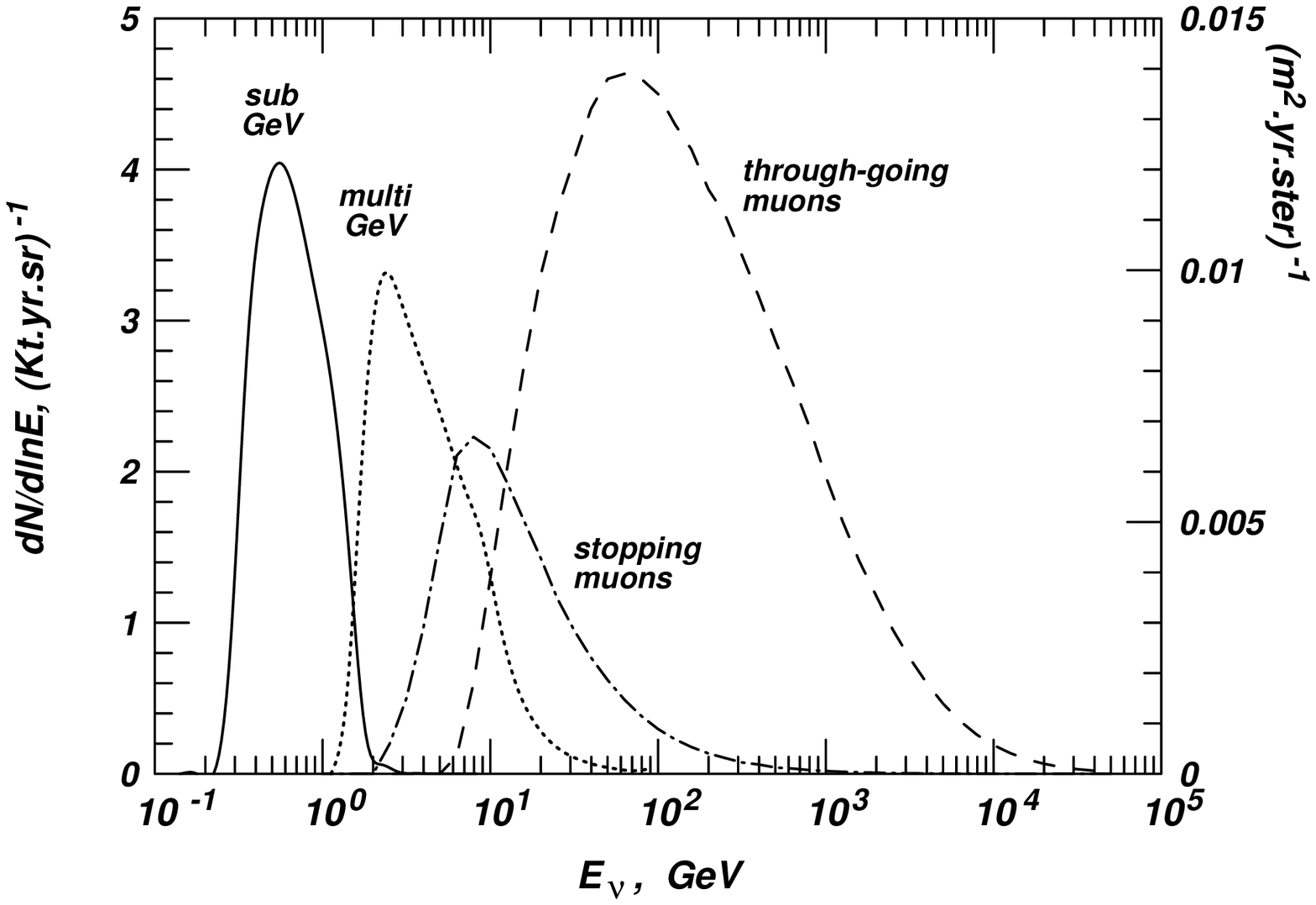}
\includegraphics[width=7cm]{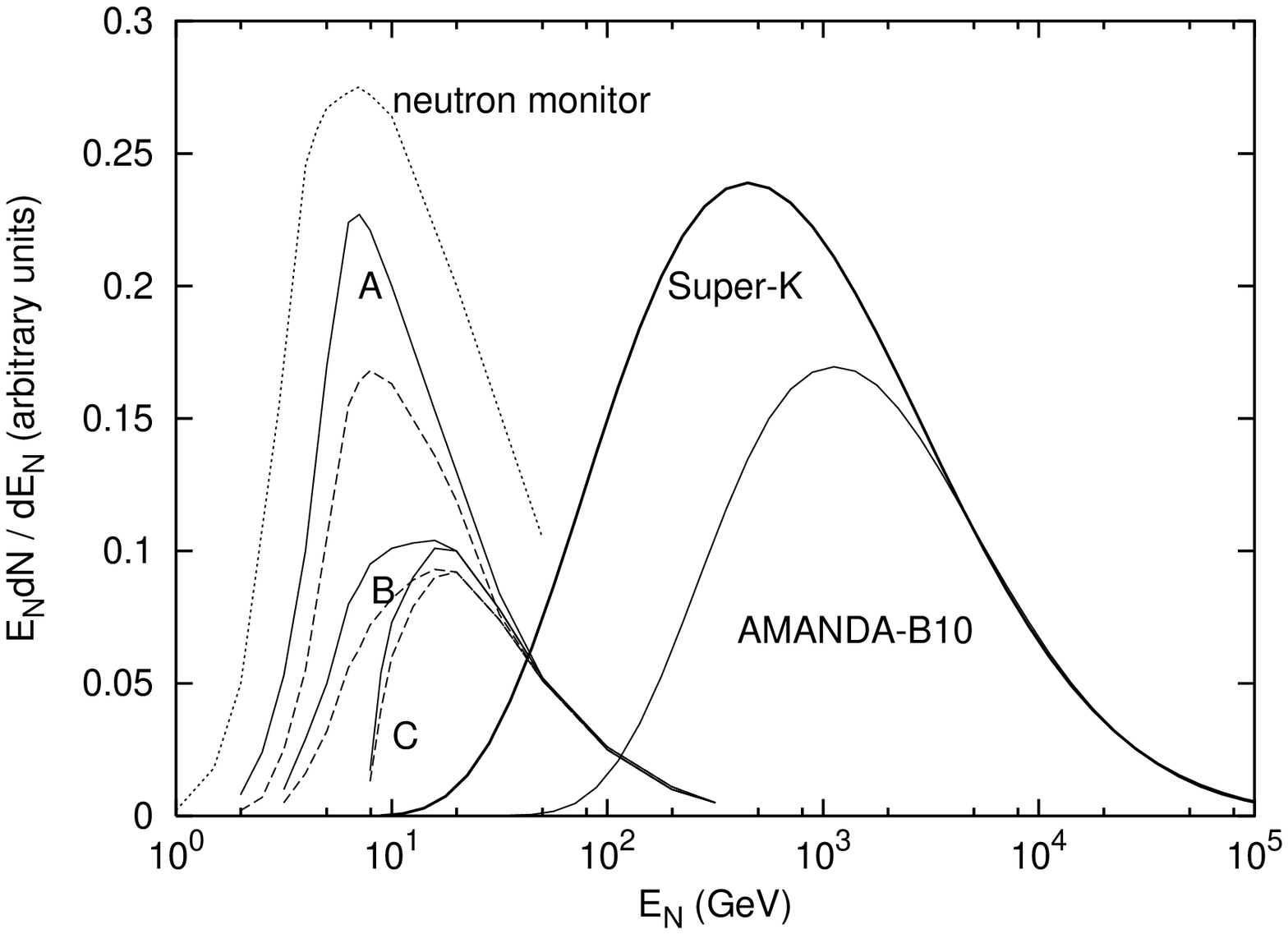}
\caption{Distribution of neutrino energies (left) and 
primary energy per nucleon (right) giving rise to
various classes of events.
}
\label{response}
\end{figure}

\section{Higher energies}

Response functions for various measurements of atmospheric neutrinos
are given in Ref.~\cite{GH} and reproduced here in
Fig.~\ref{response}.  Sub-GeV and multi-GeV
events involve neutrinos with energies below 2 GeV
and from 1 to 20 GeV respectively.  Upward stopping muons are produced
by muon neutrinos with energies in the range 3--100 GeV, while upward
throughgoing muons come from neutrinos with energies in the range
10 GeV to 10 TeV.  The relevant primary energies per nucleon are
roughly a factor of ten higher, but with rather broader distributions.
Thus the normalization of the sub-GeV neutrinos is determined almost
entirely by primaries with energies less than 100 GeV/nucleon and
hence covered by the most precise magnetic spectrometer measurements.
On the other hand, the primary response function for neutrino-induced
muons extends to 10 TeV and beyond and hence is subject to
somewhat larger uncertainty in normalization.  The difference in response
function for neutrino-induced upward muons at Super-K from that at
AMANDA arises from the higher muon energy threshold in AMANDA,
which is of order 100~GeV at the interaction vertex~\cite{AMANDAmu}.

\begin{figure}[htb]
\includegraphics[width=13cm]{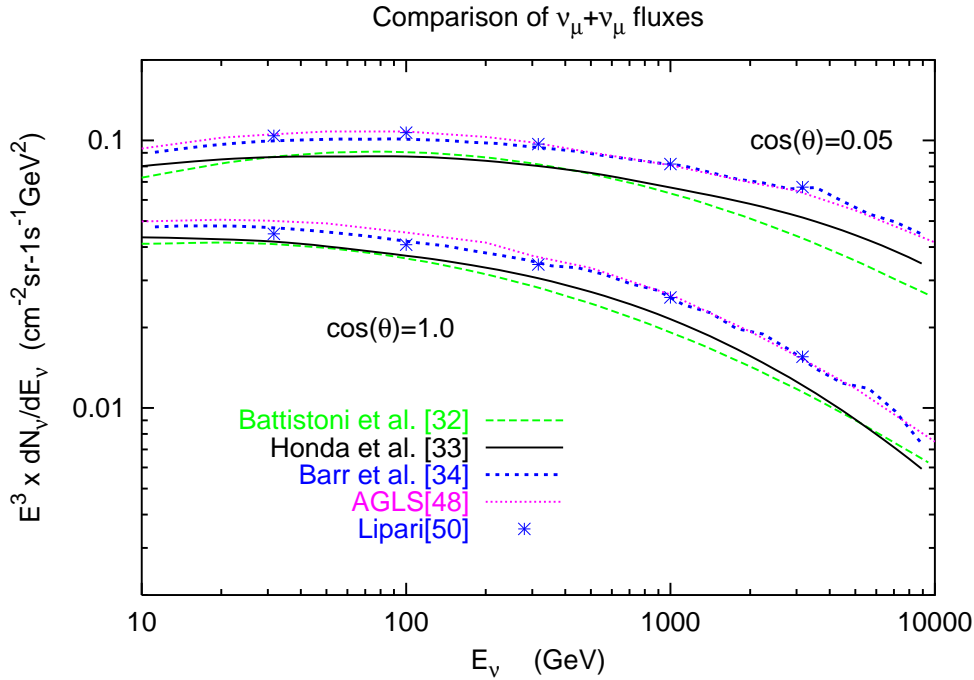}
\caption{Comparison of fluxes of muon neutrinos at high energy.}
\label{numu-hi}
\end{figure}

Again we can make use of the fact that the assumed primary spectra in
the three calculations~\cite{FLUKA,Hondaetal,Barretal} are almost the same to
illustrate differences due to the treatment of hadronic interactions.
Fig.~\ref{numu-hi} compares fluxes of muon neutrinos for these three
calculations in the energy range responsible for neutrino-induced muons.

\subsection{Analytic approximations}
At high energy, numerical integration of the analytic cascade equations
gives sufficiently accurate results to be useful for understanding the
characteristic features of the results.  The asterisks in Fig.~\ref{numu-hi}
show the numerical results of Lipari~\cite{Liparimu}, 
including the contribution from decay of muons.\footnote{The primary
spectrum of \cite{Liparimu} has the same form as Eq.~\ref{uncorrelated}
with a higher normalization.}
At high energy most
muons reach the ground before decaying and therefore
do not contribute to the neutrino flux.  For $E_\nu\,>\,100$~GeV for
example, less than 15\% of muon neutrinos are from decay of muons,
and the fraction decreases further as energy increases.

  Neglecting neutrinos
from muon decay, a simple approximation for the flux of
$\nu_\mu\,+\,\bar{\nu}_\mu$ from decay of pions and kaons is
\begin{eqnarray}
\label{analytic}
{dN_\nu\over dE_\nu}&=&{\phi_N(E_\nu)\over (1 - Z_{NN})(\gamma+1)}\left\{
\left[{Z_{N\pi}(1-r_\pi)^\gamma\over1+B_{\pi\nu}\cos\theta 
E_\nu/\epsilon_\pi}\right]\right. \\ \nonumber
&+& \left. 0.635\left[{Z_{NK}(1-r_K)^\gamma\over1+B_{K\nu}\cos\theta 
E_\nu/\epsilon_K}\right]
\right\},
\end{eqnarray}
where $\phi_N(E_\nu)$ is the primary spectrum of nucleons 
evaluated at the energy of the neutrino (see Eq.~\ref{uncorrelated}).
The constants $r_i=m_\mu^2/m_i^2$ for $i=(\pi,K)$, while the
constants $B$ depend on the hadron attenuation lengths
as well as decay kinematics.  
The critical energy
for pions is $\epsilon_\pi \approx 115$~GeV, while for kaons
$\epsilon_K\approx 850$~GeV~\cite{TKGbook}. 
 
\begin{figure}[htb]
\includegraphics[width=14cm]{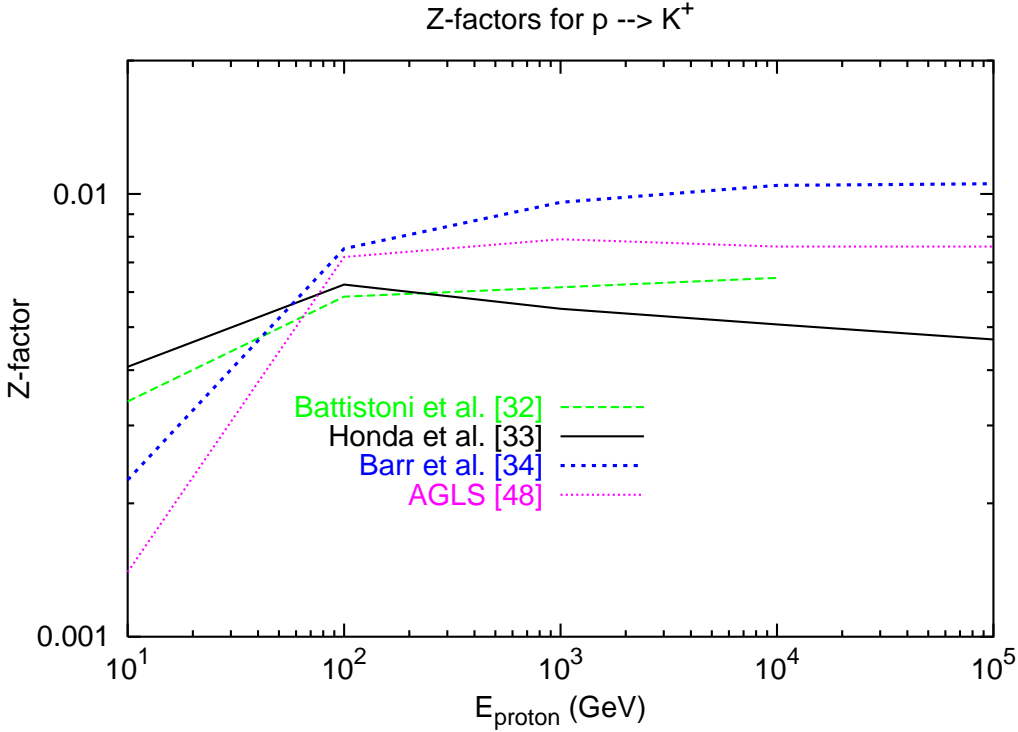}
\caption{Spectrum-weighted Z-factor for $K^+$ production
by nucleons.
}
\label{zfactors}
\end{figure}

Asymptotically, the neutrino spectrum is one power steeper
than the primary spectrum.  For fixed zenith angle $\theta$
the transition occurs first for neutrinos from pion decay
because of its lower critical energy.  As a consequence, the
relative contribution from kaons increases with energy.  Kaons
become the
dominant source of neutrinos for $E_\nu\,>\,100$~GeV.  (The onset
of kaon dominance is somewhat slower at large angle because of
the cosine factor in the denominator of the kaon term in Eq.~\ref{analytic}.)
Important differences among the calculations can be traced to
differences in treatment of kaon production, in particular
to the channel $p\,\rightarrow\,\Lambda\,K^+$.  Fig.~\ref{zfactors}
shows the spectrum-weighted moments~\cite{TKGbook} for four
calculations.

Several features of Fig.~\ref{numu-hi} can be understood with
the help of Eq.~\ref{analytic}.
\begin{itemize}
\item The general decrease of the vertical to horizontal ratio
at large energy, which appears in all
calculations, is a consequence of the factor of $\cos(\theta)$
in the denominators.
\item The primary spectrum of~\cite{Barretal} is significantly
lower than that of~\cite{AGLS} at high energy, 
as shown in Fig.~\ref{allnucleon-ratios}.  
The calculation of Ref.~\cite{Barretal} is nevertheless high, comparable
to that of Ref.~\cite{AGLS}
because the kaon production assumed in the new
calculation is larger, as shown in Fig.~\ref{zfactors}.
\item The ratio ($>1$) of the flux of Ref.~\cite{Barretal}
compared to that of Ref.~\cite{Hondaetal}, and its increase with
energy, may also be attributed to the
differences in the assumptions for kaon production as shown in Fig.~\ref{zfactors}.
\end{itemize}

\subsection{Muon fluxes}
\begin{figure}[htb]
\includegraphics[width=14cm]{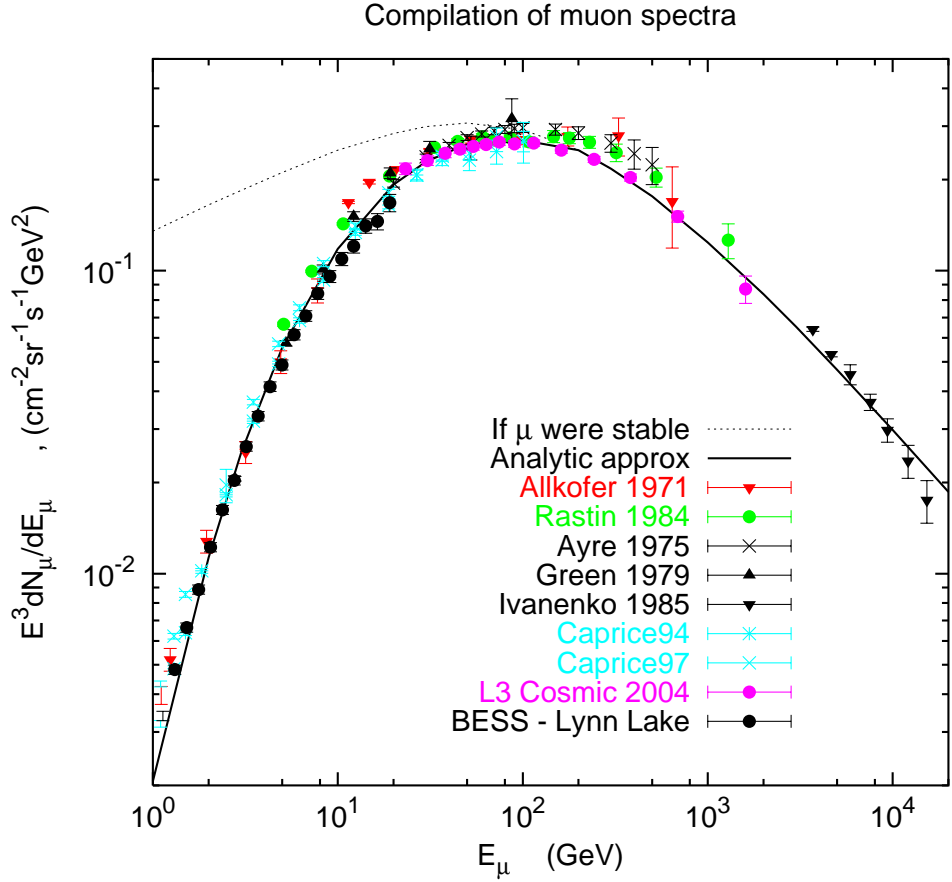}
\caption{Summary of measurements of the vertical muon intensity
at the ground.  The solid line shows an analytic 
calculation~\protect\cite{gaissermu}.
The dotted line shows the spectrum in the absence of decay and
energy loss, or equivalently the muon production spectrum integrated
over the atmosphere.
}
\label{newmuplot}
\end{figure}

Calculation of the flux of atmospheric muons follows the
same form as Eq.~\ref{convolution} with the yield of neutrinos
replaced by the yield of muons.  Because of the close relation
between muons and neutrinos, comparison of its predicted
muon intensity to
measurements is an important check on any calculation
of atmospheric muons.  Comparison to muon fluxes also
serves to check the normalization of the calculation.
Because muons lose about 2 GeV in passing through the
atmosphere, a ground level measurement probes high energies
and is therefore most relevant to multi-GeV
neutrinos and to neutrino-induced upward muons.  Because of
the kinematics of the $\pi\rightarrow\mu$ decay, however,
pion decay remains the dominant source of muons at all 
energies.  As a consequence, the leverage of the muon measurements
for controlling the neutrino flux is limited above $\sim 100$~GeV
where neutrino production is dominated by the contribution from 
kaons.

Fig.~\ref{newmuplot} contains a summary of measurements
of the spectrum of vertical muons at the ground.  All
measurements have been corrected to sea level.
At very high energy the muon
spectrum  becomes one power steeper than the parent spectrum of nucleons as a
consequence of the extra power of $1\,/\,E_\pi$ in the ratio of 
pion decay length to interaction length, which reflects
the decreasing probability of decay relative to
re-interaction for charged pions at high
energy.  For $E\,<\,\epsilon_\pi\,\approx\,115$~GeV
essentially all pions decay, and the muon production spectrum
has the same power behavior as the parent pion and grandparent nucleon
spectrum ($\alpha \approx 2.7$).  At low energy, however,
muon energy-loss and decay become important, and the muon spectrum
at the ground falls increasingly below the production spectrum.
To account for all the complications one generally resorts to 
Monte Carlo calculations.  However, analytic approximations of the
effects are also possible~\cite{Liparimu}.  Fig.~\ref{newmuplot} shows
a comparison of one such calculation~\cite{gaissermu}, which uses
as input the simple power-law primary spectrum of Eq.~\ref{uncorrelated}.
Data in Fig.~\ref{newmuplot} are from many measurements
with spectrometers at the surface \cite{Allkofer,Rastin,Ayre,Green,Capricemu,BESSmu}
and underground~\cite{L3,Ivanenko}.

The comparison of the analytic calculation with the data in Fig.~\ref{newmuplot}
leads to two remarks.  First the good agreement gives a general confirmation of
the primary spectrum at high energy.  Second, it serves to illustrate the level
of systematic differences among the various measurements.

\section{Concluding remarks}
In their recent detailed paper,~\cite{SK1489} the Super-Kamiokande group
give an extensive discussion of how the best fit oscillation
parameters are determined from a comparison of their data with
simulations based on the atmospheric neutrino flux of Ref.~\cite{Hondaetal}.
In particular, their Table VII enumerates the assumed uncertainties in
various properties of the calculated neutrino flux and the shifts in
those properties needed to obtain the best fit.  Apart from the overall
increase in normalization, the largest shifts relative to what is
assumed in Ref.~\cite{Hondaetal} concern the slope of the primary
spectrum at high energy and the K/pi ratio.  The latter is decreased
by 6\% overall, while the primary spectrum is significantly
harder as compared to the fits of Ref.~\cite{Hamburg} (from -2.71 to -2.66 above
100 GeV).  As noted in the discussion of Figs.~\ref{numu-hi} and~\ref{zfactors},
the high energy $\nu$-spectrum could also be increased by enhancing kaon production
preferentially at high energy.

The Super-K fitting
procedure uses the atmospheric neutrino data itself to correct
the calculated neutrino flux.  The neutrino data have the advantage
in principle that (apart from oscillations!) the intensity observed
at the detector integrates over the whole atmosphere without complications
of energy loss and decay that dominate the observed muon flux.
For example the required increase
in the overall normalization of 11.9\% could be interpreted as favoring
the higher normalization of Refs.~\cite{AMSp,BESS} as compared 
to that of Ref.~\cite{CAPRICE}.
The observations depend, however, on properties of neutrino interactions as well
as on the neutrino flux.  As properties of neutrino interactions
become better determined, for example from near detectors of long-baseline
experiments, this iterative procedure may become an important     way
to refine our understanding of the atmospheric neutrino flux at production.

{\bf Acknowledgments}.  I am grateful to G. Battistoni and M. Honda
for providing the Z-factors for their hadronic interaction models and
to G. Barr, G. Battistoni, M. Honda, T. Kajita, P. Lipari, S. Robbins and T. Stanev
for helpful discussions.

\end{document}